\PassOptionsToPackage{unicode}{hyperref}
\PassOptionsToPackage{hyphens}{url}
\documentclass[
  11pt]{article}
\usepackage{amsmath,amssymb}
\usepackage{iftex}
\ifPDFTeX
  \usepackage[T1]{fontenc}
  \usepackage[utf8]{inputenc}
  \usepackage{textcomp} 
\else 
  \usepackage{unicode-math} 
  \defaultfontfeatures{Scale=MatchLowercase}
  \defaultfontfeatures[\rmfamily]{Ligatures=TeX,Scale=1}
\fi
\usepackage{lmodern}
\ifPDFTeX\else
\fi
\IfFileExists{upquote.sty}{\usepackage{upquote}}{}
\IfFileExists{microtype.sty}{
  \usepackage[]{microtype}
  \UseMicrotypeSet[protrusion]{basicmath} 
}{}
\makeatletter
\@ifundefined{KOMAClassName}{
  \IfFileExists{parskip.sty}{%
    \usepackage{parskip}
  }{
    \setlength{\parindent}{0pt}
    \setlength{\parskip}{6pt plus 2pt minus 1pt}}
}{
  \KOMAoptions{parskip=half}}
\makeatother
\usepackage{xcolor}
\usepackage[margin=2.5cm]{geometry}
\usepackage{longtable,booktabs,array}
\usepackage{calc} 
\usepackage{etoolbox}
\makeatletter
\patchcmd\longtable{\par}{\if@noskipsec\mbox{}\fi\par}{}{}
\makeatother
\IfFileExists{footnotehyper.sty}{\usepackage{footnotehyper}}{\usepackage{footnote}}
\makesavenoteenv{longtable}
\usepackage{graphicx}
\makeatletter
\def\maxwidth{\ifdim\Gin@nat@width>\linewidth\linewidth\else\Gin@nat@width\fi}
\def\maxheight{\ifdim\Gin@nat@height>\textheight\textheight\else\Gin@nat@height\fi}
\makeatother
\setkeys{Gin}{width=\maxwidth,height=\maxheight,keepaspectratio}
\makeatletter
\def\fps@figure{htbp}
\makeatother
\providecommand{\tightlist}{%
  \setlength{\itemsep}{0pt}\setlength{\parskip}{0pt}}
\usepackage{newunicodechar}
\newunicodechar{Δ}{\ensuremath{\Delta}}
\newunicodechar{≥}{\ensuremath{\geq}}
\newunicodechar{≤}{\ensuremath{\leq}}
\usepackage{xurl}
\usepackage{seqsplit}
\usepackage{float}
\ifLuaTeX
  \usepackage{selnolig}  
\fi
\IfFileExists{bookmark.sty}{\usepackage{bookmark}}{\usepackage{hyperref}}
\IfFileExists{xurl.sty}{\usepackage{xurl}}{} 
\hypersetup{
  pdftitle={Is the education gap in generative AI use smaller than the internet's was at a similar level of use? Evidence from 16 European countries},
  pdfauthor={Keisuke Watanabe, Glasp Inc.},
  hidelinks,
  pdfcreator={LaTeX via pandoc}}

\title{Is the education gap in generative AI use smaller than the
internet's was at a similar level of use? Evidence from 16 European
countries}
\author{Keisuke Watanabe, Glasp Inc.}
\date{}

\begin{document}
\maketitle
\begin{abstract}
New technologies are usually adopted first by people with more formal
education. This paper asks whether the education gap in generative AI
use in 2025 is smaller or larger than the internet's gap in the same
country in the year when overall internet use was at a similar level. It
uses public Eurostat survey data on ICT usage by individuals aged 16 to
74. Each country's 2025 rate of generative AI use in the last three
months is matched to the internet year whose overall last-three-months
use was closest, within 5 percentage points, and the gap between people
with high (ISCED 5 to 8) and low (ISCED 0 to 2) completed education is
measured as a log ratio and as a percentage-point difference. Sixteen of
35 countries could be matched; most of the others were early internet
adopters. In the matched countries, the generative AI gap was smaller in
13 of 16 countries on the log ratio (median change -0.625, 95\%
bootstrap interval -0.969 to -0.291, two-sided sign test p = 0.021) and
in 15 of 16 on the percentage-point gap (median change -27.6 points,
interval -32.2 to -16.8, p = 0.0005). The point-gap result keeps its
sign in every generative-AI-versus-internet sensitivity analysis with 8
or more countries, including comparisons restricted on both sides to
ages 25 to 54 and 55 to 74, but its size is sensitive to matching that
takes education composition into account (about -10 to -11 points in 7
and 8 countries). The ratio result is fragile: it fails the decision
rule when countries matched at their first internet observation are
removed (9 of 11, p = 0.065). The analysis follows an unpublished plan
whose timing cannot be verified, so it is not presented as
pre-registered. The results describe unadjusted education contrasts in a
non-random set of countries, not an education effect, and do not
generalise to the EU as a whole.
\end{abstract}

\hypertarget{introduction}{%
\subsection{1. Introduction}\label{introduction}}

Diffusion research has long reported that earlier adopters of
innovations tend to have more years of formal education and higher
social status than later adopters (Rogers, 2003, chapter 7,
generalizations 7-3 and 7-5). For the internet, the debate in the late
1990s and 2000s turned on whether the gaps between social groups would
close as use spread (``normalization'') or persist (``stratification'')
(Norris, 2001; van Dijk and Hacker, 2003). Van Dijk and Hacker (2003)
argued that gaps in physical access could narrow through saturation
while gaps in skills and usage were likely to grow. Galácz and Smahel
(2007), comparing four countries, argued that the level of diffusion
alone does not determine the size of the divide.

Generative AI tools reached a large share of the population within a few
years of the public release of ChatGPT in November 2022. Early evidence
shows an education gradient. Bick, Blandin and Deming (NBER Working
Paper 32966; published in Management Science, 2026) find in United
States survey data that the early adoption pattern of generative AI by
education and occupation is very similar to that of the personal
computer, comparing the technologies by time since launch. Daepp and
Counts (2025) find that education is the strongest positive predictor of
county-level Bing search interest in ChatGPT in the United States, a
geographic proxy for awareness and interest rather than individual use.
Liu and Wang (2026), using website traffic data from Semrush and Google
Trends across countries up to March 2024, report that users of
generative AI tools skew young, highly educated and male compared with
general internet and Google users, and that chatbot users in particular
are more often college graduates than Google users. Their education
comparison is between the education profile of the users of two kinds of
website at one time, with user profiles inferred by Semrush from online
behaviour; it does not use survey rates of use by education group, does
not look at the internet's education gap in earlier years, and does not
match the two technologies at a similar level of overall use, which is
the comparison made here. Humlum and Vestergaard (2025) document unequal
adoption of ChatGPT among Danish workers in eleven exposed occupations.
Suárez and García-Mariñoso (2025), using a nationwide survey of actual
AI chatbot use and logit models, report that less educated people, older
people and women lag behind in use, and that use follows the pattern of
earlier digital divides. Eurostat's statistical report on the use of AI
technologies in the EU (Eurostat, 2026b, published 26 March 2026)
analyses the same 2025 survey data and reports, for the EU in 2025,
generative AI use in the last three months of 48.7\% among people with
high formal education, 25.9\% with medium and 22.4\% with no or low
formal education (p.~40, Figure 14); it does not compare these gaps with
those of earlier technologies. The version of the table used here
(updated 5 June 2026) gives 49.0\%, 25.7\% and 22.3\% for the same EU
aggregate, so the published figures appear to predate a later revision
of the table. A Eurofound working paper (Adăscăliței, 2026, p.~13)
describes the same 2025 Eurostat table across the EU27 and reports
generative AI use of 49.0\% among people with high, 25.7\% with medium
and 22.3\% with low education; it does not compare the education gap
with that of the internet or other earlier technologies at a similar
level of use, which is the comparison made here. The cross-sectional
education gradient in 2025 is therefore already documented; the
contribution here is narrower. Eurostat (2025) reported that 32.7\% of
people aged 16 to 74 in the EU used generative AI tools in 2025, and
63.8\% of those aged 16 to 24 (Eurostat, 2026a).

The comparison by time since launch, used by Bick et al., mixes two
things: how fast a technology spreads and how unequal its spread is at a
given level. A comparison at a similar level of overall use holds the
observed overall rate fixed, which removes one source of difference
between the two gaps but not others (age structure, education mix and
student share all differ between the 2000s and 2025; Section 6). The
searches made for this paper did not find a country-level comparison of
the education gap in generative AI use with the internet's gap at a
similar level of use using the same survey and recall period, but the
exact queries and screening of the earlier searches were not recorded,
so this is a provisional statement, not an established absence of prior
work (Section 9). The European survey on ICT usage is well suited to
this comparison because it has measured use in the last 3 months, with
the same education groups, for the internet since 2002 and for
generative AI since 2025. The two are distinct questionnaire items in
the same survey programme; the generative AI item was new in 2025.

This paper reports that comparison with the design, decision rule,
checks and stop conditions stated in Sections 3.2 to 3.5, which the
author took from an unpublished analysis plan, and labels every analysis
added later as exploratory. Because neither the plan's timing nor the
analysis's conformity to it can be verified from public material
(Section 3.1), the analysis is not presented as pre-registered.

\hypertarget{data}{%
\subsection{2. Data}\label{data}}

All data come from the Eurostat dissemination API (JSON-stat), fetched
on 2026-10-26 with \texttt{fetch.py} in one run that started at 06:19:44
UTC (\texttt{accessed\_utc} in the manifest): four GET requests, at
least 1.5 s apart, with no login, key, cookie or archive. Each request
was sent with Python's \texttt{urllib} and one header set by the script,
\texttt{User-Agent:\ research-replication-script/1.0\ (public\ statistics,\ non-commercial)},
from the first request of that run; urllib adds its default headers
(Host, Accept-Encoding: identity, Connection: close).
\texttt{data/MANIFEST.json} records each URL, byte count, SHA-256 hash
and the \texttt{updated} timestamp returned by Eurostat, and the raw
responses are in \texttt{data/}. This paper's run wrote its manifest,
which \texttt{fetch.py} writes only when all four requests succeed (the
console log of the fetch itself was not kept). Fetches made before this
run, for the first-pass computation, were not recorded and are not
described here; the paper's numbers come only from the fetch recorded in
the manifest. A second run of the unchanged \texttt{fetch.py} on the
same day (start 15:27:57 UTC, manifest
\texttt{data/REFETCH\_MANIFEST\_20261026.json}) received four responses
byte-identical to the deposited files, with the same \texttt{updated}
values; \texttt{verify\_extracts.py} performs this comparison and its
output is \texttt{out/verify\_extracts\_20261026.txt}. Anyone can repeat
the check while Eurostat has not revised the tables; after a revision
the hashes are expected to differ. The matched internet cells carry no
break (b) flag, which can be checked in \texttt{data/iu.json} (Section
4.3).

\begin{itemize}
\tightlist
\item
  Generative AI: dataset isoc\_ai\_iaiu, indicator I\_IUAI, ``Use of
  generative AI tools: in the last 3 months'', percent of individuals
  (PC\_IND), 2025 only, 35 countries plus aggregates. First reported in
  a Eurostat news item on 16 December 2025; the version used was last
  updated 5 June 2026 (Eurostat, isoc\_ai\_iaiu).
\item
  Internet: isoc\_ci\_ifp\_iu, indicator I\_IU3, ``Internet use: in the
  last 3 months'', percent of individuals, 2002 onward; also I\_ILT12
  (last 12 months) for one check. Last updated 17 April 2026.
\item
  Comparison technology: isoc\_ci\_ac\_i, indicator I\_IUBK (internet
  banking), percent of individuals. Last updated 17 April 2026.
\item
  Education composition (one exploratory check only): lfsa\_pgaed,
  population aged 15 to 74 by educational attainment, thousands, 2024
  and 2025. Last updated 10 September 2026.
\end{itemize}

Extract record. Each file in \texttt{data/} is the complete response to
one GET to
\path|https://ec.europa.eu/eurostat/api/dissemination/statistics/1.0/data/<table>?format=JSON&lang=EN&<filters>|,
retrieved on 2026-10-26 in the run that started at 06:19:44 UTC. That
time (\texttt{accessed\_utc} in \texttt{data/MANIFEST.json}) is taken
before the first request, so it is the start of the run, not the
retrieval time of each file; the four requests followed it, at least 1.5
s apart, and the manifest is written only after all four succeeded. The
time of each individual request was not recorded. ``Updated'' is the
\texttt{updated} field Eurostat returned in each response. SHA-256
hashes were recomputed on 2026-10-26 from the files in \texttt{data/}
and match the manifest.

{\scriptsize\setlength{\tabcolsep}{3pt}
\begin{longtable}[]{@{}
  >{\raggedright\arraybackslash}p{(\columnwidth - 10\tabcolsep) * \real{0.1667}}
  >{\raggedright\arraybackslash}p{(\columnwidth - 10\tabcolsep) * \real{0.1667}}
  >{\raggedright\arraybackslash}p{(\columnwidth - 10\tabcolsep) * \real{0.1667}}
  >{\raggedright\arraybackslash}p{(\columnwidth - 10\tabcolsep) * \real{0.1667}}
  >{\raggedright\arraybackslash}p{(\columnwidth - 10\tabcolsep) * \real{0.1667}}
  >{\raggedright\arraybackslash}p{(\columnwidth - 10\tabcolsep) * \real{0.1667}}@{}}
\toprule\noalign{}
\begin{minipage}[b]{\linewidth}\raggedright
File
\end{minipage} & \begin{minipage}[b]{\linewidth}\raggedright
Source table
\end{minipage} & \begin{minipage}[b]{\linewidth}\raggedright
Filters (besides format=JSON, lang=EN)
\end{minipage} & \begin{minipage}[b]{\linewidth}\raggedright
Updated (returned)
\end{minipage} & \begin{minipage}[b]{\linewidth}\raggedright
Bytes
\end{minipage} & \begin{minipage}[b]{\linewidth}\raggedright
SHA-256
\end{minipage} \\
\midrule\noalign{}
\endhead
\bottomrule\noalign{}
\endlastfoot
ai.json & isoc\_ai\_iaiu & unit=PC\_IND, indic\_is=I\_IUAI &
2026-06-05T11:00:00+0200 & 56,214 &
\texttt{\seqsplit{0c5786da65c759131af888ad0320eb39d7a52fba2d1daa9710f059d7b506094b}} \\
iu.json & isoc\_ci\_ifp\_iu & unit=PC\_IND, indic\_is=I\_IU3,
indic\_is=I\_ILT12 & 2026-04-17T11:00:00+0200 & 2,310,306 &
\texttt{\seqsplit{9370baa933e4475abab13f95e9047348c7e597481700a10f37a6b75ad0951683}} \\
bk.json & isoc\_ci\_ac\_i & unit=PC\_IND, indic\_is=I\_IUBK &
2026-04-17T11:00:00+0200 & 1,083,437 &
\texttt{\seqsplit{f0da47ee994e4e941d6f9cef4449d1bc8fc9091ecad8fc1d54a6990ad8ed0506}} \\
pg.json & lfsa\_pgaed & unit=THS\_PER, sex=T, age=Y15-74, isced11=ED0-2,
ED3\_4, ED5-8, time=2024, 2025 & 2026-09-10T23:00:00+0200 & 7,389 &
\texttt{\seqsplit{914941e5384317229fdcb3123d0d5497ecca7b654b36ab047fcfbb35d8fa36c6}} \\
\end{longtable}
}

The earlier fetch or fetches used before this paper (for the first-pass
computation above, and for an earlier revision of the analysis outputs
made the day before this fetch) were not saved; their retrieval times
and hashes were not recorded and cannot be given. The only links to this
fetch are outputs: the three output files of that earlier revision are
reproduced byte for byte from these extracts (Section 8), and the M1 and
F1 to F7 values printed in the first results draft are reproduced at
their printed precision. Eurostat may revise the tables; a later
download can be compared with these hashes.

The ICT survey covers individuals aged 16 to 74. Education is the
highest level of education completed, so people still in education are
classified by what they have completed so far (Eurostat metadata,
isoc\_i\_esms). Low education is ISCED 0 to 2 (code I0\_2), medium is
ISCED 3 to 4 (I3\_4), high is ISCED 5 to 8 (I5\_8). All internet years
used in the matched comparison fall before 2014 and use ISCED 1997; the
2025 generative AI values use ISCED 2011.

Eurostat statistical data may be reused with acknowledgement of the
source under Commission Decision 2011/833/EU (Eurostat copyright
notice).

\hypertarget{design}{%
\subsection{3. Design}\label{design}}

\hypertarget{analysis-plan-and-why-it-is-not-a-pre-registration}{%
\subsubsection{3.1 Analysis plan and why it is not a
pre-registration}\label{analysis-plan-and-why-it-is-not-a-pre-registration}}

The design was written in an analysis plan (in Japanese) that is not
public. The code, data and outputs are available as ancillary files with
this arXiv submission (folder \texttt{anc/}); \texttt{anc/README.md}
gives the commands to rerun them. According to the author's version
history, which is not public, the plan was recorded on 2026-10-25 about
six minutes before the first results draft, which reports the M1 result
(16 countries, median -0.625 and -27.6), and the plan has not changed
since. Readers cannot check this record, and nothing in this paper
depends on it. No record exists of when the first-pass script was run.

This record shows the order in which the two files were recorded, not
the order in which the plan was written and the computation was run. The
timestamps are set by the author's machine and are not public; the
first-pass analysis script was not kept with the plan and its run time
is not recorded; and six minutes is short enough that the computation
may have been run before the plan was recorded. The author had also seen
some values beforehand (EU-wide 2025 generative AI rates by education;
internet rates by education for Greece 2007, Italy 2005 and Portugal
2005) and expected the ``smaller'' direction (plan, Section 0). For
these reasons the analysis is presented as not pre-registered.
``Planned'' below means that the author states the analysis was
specified in this plan; the statement cannot be verified, and it does
not carry the evidential weight of a pre-registration. Analyses added
after the first results are labelled exploratory. The distinction
between planned and exploratory analyses is therefore the author's
classification, and readers who do not accept it can treat every
analysis in this paper as exploratory; all of them are reported.

\hypertarget{primary-analysis-m1}{%
\subsubsection{3.2 Primary analysis (M1)}\label{primary-analysis-m1}}

For each country c:

\begin{enumerate}
\def\labelenumi{\arabic{enumi}.}
\tightlist
\item
  T\_c is the 2025 generative AI rate for all individuals aged 16 to 74.
\item
  The matched internet year y*\_c is the year in which internet use
  (I\_IU3) for all individuals, low education and high education are all
  present without a ``low reliability'' (u) flag, and whose overall rate
  is closest to T\_c.~Ties go to the earlier year. The country counts as
  matched only if the distance is at most 5.0 points. A country is
  excluded if any of its three generative AI cells has a u flag.
\item
  With p\_H and p\_L the high- and low-education rates, the gap is
  measured as the log ratio ln R = ln(p\_H / p\_L) and the difference D
  = p\_H - p\_L (points). The log-odds difference L = logit(p\_H) -
  logit(p\_L) is secondary.
\item
  The country-level change is Δ = gap(generative AI, 2025) -
  gap(internet, y*\_c). A negative Δ means the generative AI gap is
  smaller.
\end{enumerate}

Across matched countries, the summary is the median of Δ, a two-sided
exact sign test with countries as units, and a percentile bootstrap 95\%
interval for the median (10,000 resamples of countries with replacement;
Python \texttt{random.Random} seeded with 20261026; median of an even
count is the mean of the two middle values; interval endpoints are the
sorted resample medians at zero-based positions 250 and 9749). ln R and
D are both primary.

\hypertarget{decision-rule-planned}{%
\subsubsection{3.3 Decision rule
(planned)}\label{decision-rule-planned}}

\begin{itemize}
\tightlist
\item
  Both medians negative and both sign tests p \textless{} 0.05:
  ``smaller''.
\item
  Both medians positive and both p \textless{} 0.05: ``larger''.
\item
  Medians of opposite sign: ``depends on the measure''.
\item
  Same sign but at least one p ≥ 0.05: ``not distinguishable''.
\end{itemize}

\hypertarget{falsification-and-robustness-checks-planned}{%
\subsubsection{3.4 Falsification and robustness checks
(planned)}\label{falsification-and-robustness-checks-planned}}

\begin{itemize}
\tightlist
\item
  F1: linear interpolation between the two internet years bracketing
  T\_c.
\item
  F2: match on the medium-education rate instead of the overall rate.
\item
  F3: internet use in the last 12 months.
\item
  F4: internet banking, matched to T\_c in the same way, compared with
  the internet (Δ = banking gap - internet gap), to ask whether any
  difference is specific to generative AI or common to technologies that
  spread after the internet.
\item
  F5: ages 25 to 64 on both sides (internet cells from 2009 only).
\item
  F6: EU member states only.
\item
  F7: drop countries matched at their first internet observation.
\end{itemize}

\hypertarget{stop-conditions-planned}{%
\subsubsection{3.5 Stop conditions
(planned)}\label{stop-conditions-planned}}

\begin{enumerate}
\def\labelenumi{\arabic{enumi}.}
\tightlist
\item
  Fewer than 8 matched countries: no conclusion.
\item
  Left censoring: if more than half of matched countries were matched at
  their first internet observation with that year's overall rate already
  above T\_c, the claim ``smaller'' may not be made.
\item
  Age composition: if F5 has 8 or more countries and both measures
  reverse sign, weaken the headline. With fewer than 8, state that the
  age-matched comparison cannot be made with these data.
\item
  Specificity: if internet banking (compared with the internet) shows
  the same direction as generative AI with at least half the magnitude,
  describe the result as common to later technologies rather than
  specific to generative AI.
\item
  Breaks in series: if more than half of the matched internet cells
  carry a break (b) flag, say so first.
\end{enumerate}

\hypertarget{results}{%
\subsection{4. Results}\label{results}}

\hypertarget{matching}{%
\subsubsection{4.1 Matching}\label{matching}}

Sixteen of 35 countries matched: Austria, Bulgaria, Cyprus, Czechia,
Estonia, Greece, Spain, Hungary, Lithuania, Latvia, Malta, Portugal,
Romania and Slovenia (EU members), and North Macedonia and Turkey
(candidate countries). Matched internet years range from 2003 to 2010,
and match residuals (internet overall minus T\_c) range from -3.3 to
+3.8 points (Table 1).

Thirteen EU members did not match. In 12 of them (Belgium, Germany,
Denmark, Finland, France, Croatia, Italy, Luxembourg, the Netherlands,
Poland, Sweden, Slovakia), the first observed internet rate with
complete education cells was more than 5 points above the 2025
generative AI rate. Ireland had internet years below and above its
target but none within 5 points. Of the non-EU countries, Norway,
Switzerland, Albania, Bosnia and Herzegovina, Serbia and Kosovo also
started above their target. The matched set therefore systematically
excludes countries where the internet spread early, and the results
apply to these 16 countries only.

Table 1. Primary match (M1). Rates in percent of individuals aged 16 to
74.

{\scriptsize\setlength{\tabcolsep}{3pt}
\begin{longtable}[]{@{}
  >{\raggedright\arraybackslash}p{(\columnwidth - 18\tabcolsep) * \real{0.1000}}
  >{\raggedright\arraybackslash}p{(\columnwidth - 18\tabcolsep) * \real{0.1000}}
  >{\raggedright\arraybackslash}p{(\columnwidth - 18\tabcolsep) * \real{0.1000}}
  >{\raggedright\arraybackslash}p{(\columnwidth - 18\tabcolsep) * \real{0.1000}}
  >{\raggedright\arraybackslash}p{(\columnwidth - 18\tabcolsep) * \real{0.1000}}
  >{\raggedright\arraybackslash}p{(\columnwidth - 18\tabcolsep) * \real{0.1000}}
  >{\raggedright\arraybackslash}p{(\columnwidth - 18\tabcolsep) * \real{0.1000}}
  >{\raggedright\arraybackslash}p{(\columnwidth - 18\tabcolsep) * \real{0.1000}}
  >{\raggedright\arraybackslash}p{(\columnwidth - 18\tabcolsep) * \real{0.1000}}
  >{\raggedright\arraybackslash}p{(\columnwidth - 18\tabcolsep) * \real{0.1000}}@{}}
\toprule\noalign{}
\begin{minipage}[b]{\linewidth}\raggedright
Country
\end{minipage} & \begin{minipage}[b]{\linewidth}\raggedright
GenAI 2025 all
\end{minipage} & \begin{minipage}[b]{\linewidth}\raggedright
GenAI low
\end{minipage} & \begin{minipage}[b]{\linewidth}\raggedright
GenAI high
\end{minipage} & \begin{minipage}[b]{\linewidth}\raggedright
Internet year
\end{minipage} & \begin{minipage}[b]{\linewidth}\raggedright
Internet all
\end{minipage} & \begin{minipage}[b]{\linewidth}\raggedright
Internet low
\end{minipage} & \begin{minipage}[b]{\linewidth}\raggedright
Internet high
\end{minipage} & \begin{minipage}[b]{\linewidth}\raggedright
Δ ln R
\end{minipage} & \begin{minipage}[b]{\linewidth}\raggedright
Δ D (points)
\end{minipage} \\
\midrule\noalign{}
\endhead
\bottomrule\noalign{}
\endlastfoot
AT & 39.4 & 28.0 & 54.1 & 2003 & 41.0 & 22.3 & 78.9 & -0.605 & -30.5 \\
BG & 22.5 & 13.6 & 41.3 & 2006 & 24.4 & 11.0 & 57.8 & -0.547 & -19.1 \\
CY & 44.2 & 19.9 & 64.6 & 2009 & 47.9 & 20.6 & 78.7 & -0.162 & -13.4 \\
CZ & 35.4 & 47.2 & 55.8 & 2005 & 32.1 & 25.7 & 72.7 & -0.872 & -38.4 \\
EE & 46.6 & 47.6 & 54.7 & 2004 & 50.5 & 42.3 & 69.3 & -0.355 & -19.9 \\
EL & 44.1 & 18.9 & 66.4 & 2010 & 44.4 & 16.1 & 75.9 & -0.291 & -12.2 \\
ES & 37.9 & 20.2 & 50.6 & 2003 & 36.6 & 14.9 & 71.2 & -0.644 & -25.8 \\
HU & 29.6 & 21.9 & 45.1 & 2004 & 27.7 & 12.5 & 68.0 & -0.969 & -32.2 \\
LT & 36.9 & 44.0 & 53.9 & 2005 & 34.3 & 28.4 & 68.6 & -0.679 & -30.3 \\
LV & 33.4 & 42.4 & 46.9 & 2004 & 33.1 & 19.2 & 63.8 & -1.099 & -40.0 \\
MK & 22.0 & 10.5 & 43.8 & 2004 & 21.1 & 10.8 & 43.9 & +0.024 & +0.2 \\
MT & 46.5 & 14.9 & 71.5 & 2007 & 45.3 & 28.4 & 91.6 & +0.397 & -6.6 \\
PT & 38.7 & 14.6 & 63.4 & 2007 & 39.6 & 23.9 & 89.5 & +0.147 & -16.8 \\
RO & 17.8 & 10.0 & 41.0 & 2006 & 20.9 & 1.7 & 65.8 & -2.263 & -33.0 \\
SI & 37.6 & 31.3 & 59.0 & 2004 & 36.9 & 16.4 & 84.4 & -1.008 & -40.4 \\
TR & 17.2 & 6.4 & 35.8 & 2005 & 13.9 & 3.8 & 62.6 & -1.072 & -29.4 \\
\end{longtable}
}

Across the 16 countries, the median high-minus-low gap was 28.6 points
for generative AI and 56.4 points for the internet at the matched year;
the median log ratio was 1.016 (a ratio of about 2.8) and 1.374 (about
3.9). With 16 countries, each median is the mean of the eighth and ninth
sorted values: for the generative AI log ratio these are Spain (0.920)
and Bulgaria (1.113), for the internet log ratio Cyprus (1.341) and
North Macedonia (1.407) (\texttt{describe\_m1.py}, output
\texttt{out/m1\_descriptive.txt}).

In four countries (Czechia, Estonia, Lithuania, Latvia), the 2025
generative AI rate among people with low completed education is above
the overall rate. This is consistent with the low-education group at
ages 16 to 74 containing many young people who have not yet completed
upper secondary education; it is a descriptive observation, not a tested
explanation.

\hypertarget{primary-result-and-planned-checks}{%
\subsubsection{4.2 Primary result and planned
checks}\label{primary-result-and-planned-checks}}

Table 2. Planned analyses (M1, F1 to F7). Median of Δ with 95\%
bootstrap interval; counts of countries with a negative and a positive
Δ; two-sided exact sign test. For M1, F1 to F3 and F5 to F7, Δ is the
generative AI gap minus the internet gap, so negative means a smaller
generative AI gap. For F4, Δ is the internet banking gap minus the
internet gap; negative means a smaller banking gap, and F4 says nothing
directly about generative AI.

{\scriptsize\setlength{\tabcolsep}{3pt}
\begin{longtable}[]{@{}
  >{\raggedright\arraybackslash}p{(\columnwidth - 16\tabcolsep) * \real{0.1111}}
  >{\raggedright\arraybackslash}p{(\columnwidth - 16\tabcolsep) * \real{0.1111}}
  >{\raggedright\arraybackslash}p{(\columnwidth - 16\tabcolsep) * \real{0.1111}}
  >{\raggedright\arraybackslash}p{(\columnwidth - 16\tabcolsep) * \real{0.1111}}
  >{\raggedright\arraybackslash}p{(\columnwidth - 16\tabcolsep) * \real{0.1111}}
  >{\raggedright\arraybackslash}p{(\columnwidth - 16\tabcolsep) * \real{0.1111}}
  >{\raggedright\arraybackslash}p{(\columnwidth - 16\tabcolsep) * \real{0.1111}}
  >{\raggedright\arraybackslash}p{(\columnwidth - 16\tabcolsep) * \real{0.1111}}
  >{\raggedright\arraybackslash}p{(\columnwidth - 16\tabcolsep) * \real{0.1111}}@{}}
\toprule\noalign{}
\begin{minipage}[b]{\linewidth}\raggedright
Analysis
\end{minipage} & \begin{minipage}[b]{\linewidth}\raggedright
n
\end{minipage} & \begin{minipage}[b]{\linewidth}\raggedright
Δ ln R median (95\% interval)
\end{minipage} & \begin{minipage}[b]{\linewidth}\raggedright
negative / positive
\end{minipage} & \begin{minipage}[b]{\linewidth}\raggedright
p
\end{minipage} & \begin{minipage}[b]{\linewidth}\raggedright
Δ D median, points (95\% interval)
\end{minipage} & \begin{minipage}[b]{\linewidth}\raggedright
negative / positive
\end{minipage} & \begin{minipage}[b]{\linewidth}\raggedright
p
\end{minipage} & \begin{minipage}[b]{\linewidth}\raggedright
Rule
\end{minipage} \\
\midrule\noalign{}
\endhead
\bottomrule\noalign{}
\endlastfoot
M1 primary & 16 & -0.625 (-0.969 to -0.291) & 13 / 3 & 0.021 & -27.6
(-32.2 to -16.8) & 15 / 1 & 0.0005 & smaller \\
F1 interpolation & 16 & -0.652 (-0.993 to -0.305) & 13 / 3 & 0.021 &
-27.4 (-34.4 to -15.4) & 15 / 1 & 0.0005 & smaller \\
F2 match on medium-education rate & 8 & -0.980 (-1.302 to -0.542) & 8 /
0 & 0.008 & -10.0 (-41.6 to -2.3) & 8 / 0 & 0.008 & smaller \\
F3 internet, last 12 months & 12 & -0.628 (-0.896 to -0.370) & 11 / 1 &
0.006 & -30.9 (-36.2 to -15.7) & 12 / 0 & 0.0005 & smaller \\
F4 banking minus internet & 15 & +0.331 (+0.021 to +1.252) & 3 / 12 &
0.035 & -1.5 (-8.0 to +9.3) & 8 / 7 & 1.0 & depends on measure \\
F5 ages 25-64, both sides & 2 & -0.494 (-0.616 to -0.373) & 2 / 0 & 0.5
& -15.9 (-16.6 to -15.2) & 2 / 0 & 0.5 & fewer than 8 \\
F6 EU members only & 14 & -0.625 (-0.940 to -0.291) & 12 / 2 & 0.013 &
-28.1 (-33.0 to -16.8) & 14 / 0 & 0.0001 & smaller \\
F7 drop first-year matches & 11 & -0.605 (-0.872 to -0.162) & 9 / 2 &
0.065 & -25.8 (-30.5 to -13.4) & 11 / 0 & 0.001 & not distinguishable \\
\end{longtable}
}

Under the plan's decision rule, the primary result is ``smaller'': the
generative AI gap in 2025 is smaller than the internet's gap at a
similar overall level of use, on both the ratio and the point measure.
The secondary log-odds difference agrees (median -1.28, 95\% interval
-1.64 to -0.84, 15 of 16 smaller, p = 0.0005).

The planned checks qualify this. F1, F3 and F6 agree. F2 agrees on both
measures, but with only 8 countries and a much smaller point change
(-10.0 instead of -27.6), so the size of the point gap depends on how
the level is matched. F7, which removes the five countries matched at
their first internet observation (Estonia, Hungary, Latvia, North
Macedonia, Slovenia), keeps the point result (11 of 11) but the ratio
result drops to 9 of 11 (p = 0.065), which the plan's rule classes as
not distinguishable. The plan defined F7 as a falsification check that
does not change the primary interpretation, and the primary verdict is
reported as the plan states; the ratio part of that verdict nonetheless
fails this check, and that failure is reported as a failure.

\hypertarget{stop-conditions}{%
\subsubsection{4.3 Stop conditions}\label{stop-conditions}}

\begin{enumerate}
\def\labelenumi{\arabic{enumi}.}
\tightlist
\item
  Too few countries: does not apply (16 matched).
\item
  Left censoring, as planned: one of 16 matched countries (Estonia) was
  matched at its first observation with an internet rate already above
  T\_c.~The condition (more than half) does not apply. This rule only
  sees matched countries; the 12 EU members that failed to match because
  their internet series started above target are a selection problem for
  generalisation (Section 4.1), not for the internal comparison.
\item
  Age composition: F5 has 2 countries, so this condition applies in its
  fewer-than-8 branch. As planned, the age-matched comparison at ages 25
  to 64 cannot be made with these data. This condition does not prevent
  the primary statement.
\item
  Specificity: internet banking minus internet goes the opposite way on
  the ratio (median +0.331, 12 of 15 positive, p = 0.035) and shows no
  distinguishable difference in points (median -1.5, 95\% interval -8.0
  to +9.3, 8 of 15 negative, p = 1.0). The condition does not apply. On
  the ratio, banking does not share the smaller gap; on points, the data
  give no evidence that banking has a smaller gap than the internet,
  which is weaker than showing that it does not.
\item
  Breaks in series: none of the 48 matched internet cells (16 countries,
  three cells each) carries a b flag. \texttt{out/flag\_audit.csv},
  written by \texttt{audit\_flags.py} from the raw extracts in
  \texttt{data/}, lists the status flag of every cell that enters the
  reported values of M1, F2 and R2 (generative AI and internet total,
  low, medium where used, and high; for R2 also the three LFS cells of
  the weight year actually used, 2025 for all seven countries) and of
  the student and 16-24 comparison: 294 cells. None carries a b flag;
  one carries a u flag (North Macedonia, generative AI 2025 among
  students, 53.32, excluded); one is missing (Spain, internet use among
  students in 2003). That file covers the selected cells only. Because
  the matching step skips u-flagged cells before it picks the closest
  year, cells of years that were examined but not selected can also
  affect the result. A second file,
  \texttt{out/flag\_audit\_candidates.csv}, written by the same script,
  lists every cell that the matching step of M1, F2 and R2 can examine:
  for all 35 countries, the generative AI total, low, medium and high
  cells for 2025, the internet total, low, medium and high cells in
  every year present in the extract, and the LFS weight cells for 2024
  and 2025 (3,186 cells). Five of these carry a u flag (Denmark 2015,
  low, medium and high; Ireland 2003, low; Slovakia 2009, low), and all
  five have no value, so they are skipped as missing whether or not the
  flag rule applies; no cell with a value carries a u flag. The flag
  rule therefore did not change which year was selected in M1, F2 or R2.
  Cells with a b flag do occur among candidate years (internet: 8 total,
  11 low, 12 medium and 12 high cells; LFS: 6 cells);
  \texttt{analysis.py} does not exclude b-flagged cells, and none of
  them is among the selected cells listed above. Cells of the other
  analyses (F1, F3 to F7, R1 and R3 to R5) are not listed in either
  file; \texttt{analysis.py} treats u-flagged cells as missing in all
  analyses.
\end{enumerate}

\hypertarget{exploratory-analyses-not-in-the-plan}{%
\subsubsection{4.4 Exploratory analyses (not in the
plan)}\label{exploratory-analyses-not-in-the-plan}}

These analyses were added after the primary result was known. They are
exploratory.

Table 3. Exploratory analyses.

{\scriptsize\setlength{\tabcolsep}{3pt}
\begin{longtable}[]{@{}
  >{\raggedright\arraybackslash}p{(\columnwidth - 14\tabcolsep) * \real{0.1250}}
  >{\raggedright\arraybackslash}p{(\columnwidth - 14\tabcolsep) * \real{0.1250}}
  >{\raggedright\arraybackslash}p{(\columnwidth - 14\tabcolsep) * \real{0.1250}}
  >{\raggedright\arraybackslash}p{(\columnwidth - 14\tabcolsep) * \real{0.1250}}
  >{\raggedright\arraybackslash}p{(\columnwidth - 14\tabcolsep) * \real{0.1250}}
  >{\raggedright\arraybackslash}p{(\columnwidth - 14\tabcolsep) * \real{0.1250}}
  >{\raggedright\arraybackslash}p{(\columnwidth - 14\tabcolsep) * \real{0.1250}}
  >{\raggedright\arraybackslash}p{(\columnwidth - 14\tabcolsep) * \real{0.1250}}@{}}
\toprule\noalign{}
\begin{minipage}[b]{\linewidth}\raggedright
Analysis
\end{minipage} & \begin{minipage}[b]{\linewidth}\raggedright
n
\end{minipage} & \begin{minipage}[b]{\linewidth}\raggedright
Δ ln R median (95\% interval)
\end{minipage} & \begin{minipage}[b]{\linewidth}\raggedright
smaller / larger
\end{minipage} & \begin{minipage}[b]{\linewidth}\raggedright
p
\end{minipage} & \begin{minipage}[b]{\linewidth}\raggedright
Δ D median (95\% interval)
\end{minipage} & \begin{minipage}[b]{\linewidth}\raggedright
smaller / larger
\end{minipage} & \begin{minipage}[b]{\linewidth}\raggedright
p
\end{minipage} \\
\midrule\noalign{}
\endhead
\bottomrule\noalign{}
\endlastfoot
R1 tolerance 2.5 points & 10 & -0.576 (-0.969 to +0.024) & 7 / 3 & 0.34
& -22.5 (-35.3 to -11.7) & 9 / 1 & 0.021 \\
R2 composition-standardised match & 7 & -0.572 (-1.426 to -0.513) & 6 /
1 & 0.125 & -11.5 (-41.6 to -2.3) & 6 / 1 & 0.125 \\
R3a ages 25-54 both sides, matched on 25-54 rate & 12 & -0.322 (-1.004
to -0.074) & 10 / 2 & 0.039 & -21.7 (-31.0 to -11.7) & 12 / 0 &
0.0005 \\
R3b ages 25-54 both sides, at the M1 year & 11 & -0.622 (-1.391 to
-0.027) & 10 / 1 & 0.012 & -21.5 (-27.5 to -10.0) & 11 / 0 & 0.001 \\
R3c as R3a, drop first-year matches & 8 & -0.290 (-1.391 to +0.097) & 6
/ 2 & 0.29 & -19.3 (-22.5 to -8.7) & 8 / 0 & 0.008 \\
R4a ages 55-74 both sides, matched on 55-74 rate & 18 & -0.397 (-0.932
to -0.061) & 13 / 5 & 0.096 & -19.9 (-24.3 to -15.8) & 18 / 0 &
\textless{} 0.0001 \\
R4b ages 55-74 both sides, at the M1 year & 6 & -1.526 (-2.570 to
-0.090) & 5 / 1 & 0.22 & -14.9 (-20.5 to -10.9) & 6 / 0 & 0.031 \\
R5a EU members, drop first-year matches & 10 & -0.576 (-0.758 to -0.162)
& 8 / 2 & 0.11 & -22.5 (-31.6 to -13.4) & 10 / 0 & 0.002 \\
R5b EU members, 25-54 at the M1 year & 10 & -0.451 (-1.391 to -0.027) &
9 / 1 & 0.021 & -20.6 (-25.8 to -10.0) & 10 / 0 & 0.002 \\
R5c EU members, 25-54 matched on 25-54 rate & 12 & -0.322 (-1.004 to
-0.074) & 10 / 2 & 0.039 & -21.7 (-31.0 to -11.7) & 12 / 0 & 0.0005 \\
\end{longtable}
}

\begin{itemize}
\tightlist
\item
  R2 reweights each internet year's low-, medium- and high-education
  rates to the 2025 education mix of the population aged 15 to 74
  (Labour Force Survey, lfsa\_pgaed) and matches T\_c to that
  standardised rate. The question it addresses is that the same overall
  rate can hide different group rates when the education mix has
  shifted. R2 is an approximate check, not a decomposition of the M1
  change: it uses population weights for ages 15 to 74 against ICT rates
  for ages 16 to 74; it changes the matched years and the countries (7
  countries match: Bulgaria, Cyprus, Czechia, Greece, Ireland, North
  Macedonia and Romania, of which Ireland is not in M1); and it still
  compares unstandardised high-minus-low gaps. The code takes the
  education weights from one year for all three groups (2025, else
  2024); all 7 matched countries have 2025 values for all three groups,
  so no country mixes years, and no LFS cell used carries a u flag. In
  R2 the median point change is -11.5 (6 of 7 negative, p = 0.125),
  close to F2 (-10.0, 8 countries). The size of the point-gap difference
  is therefore sensitive to matching that takes education composition
  into account. These checks do not show how much of the M1 change is
  due to composition.
\item
  R3 and R4 restrict both technologies to the same age band. Ages 25 to
  54 exclude most students and the oldest group on both sides; the
  internet cells for these bands exist from 2004. The point change stays
  negative in every country in R3a, R3b, R3c, R4a and R4b. The ratio
  change stays negative in the median but meets p \textless{} 0.05 only
  in R3a and R3b. Low-education rates in some countries at these ages
  are close to zero in the internet era, which makes ratios unstable.
\item
  Leave-one-country-out on the M1 ratio: the sign-test p ranges from
  0.007 to 0.035 across the 16 omissions.
\item
  Students: the median internet use rate among students in the matched
  year was 85.7\% (15 countries with data), and the median 2025
  generative AI rate among students was 85.5\% (15 countries; North
  Macedonia's value carries a u flag and is excluded under the flag rule
  in Section 8). Heavy student use is present on both sides of the
  comparison.
\item
  An earlier draft changed the age band to 25 to 64 on the generative AI
  side only, keeping the internet at 16 to 74. That asymmetric
  comparison reversed the ratio result (5 of 16 smaller) and is not a
  valid test of age composition; it is superseded by R3 and R4.
\end{itemize}

Figure 1 (\texttt{figs/fig1\_trajectories.png}) shows, for each matched
country, the internet trajectory of the high/low ratio against overall
use from 2002 to 2012, with the matched year ringed and the 2025
generative AI point overlaid. Figure 2 (\texttt{figs/fig2\_forest.png})
shows the median Δ and its interval for every median-and-interval
analysis in Tables 2 and 3 except F4 (R6, the leave-one-country-out
range, has no single median and interval and is not plotted); blue marks
the planned analyses (M1, F1 to F7, from the analysis plan, not
pre-registered) and orange with an asterisk marks the exploratory ones.

\begin{figure}
\centering
\includegraphics[width=1\textwidth,height=\textheight]{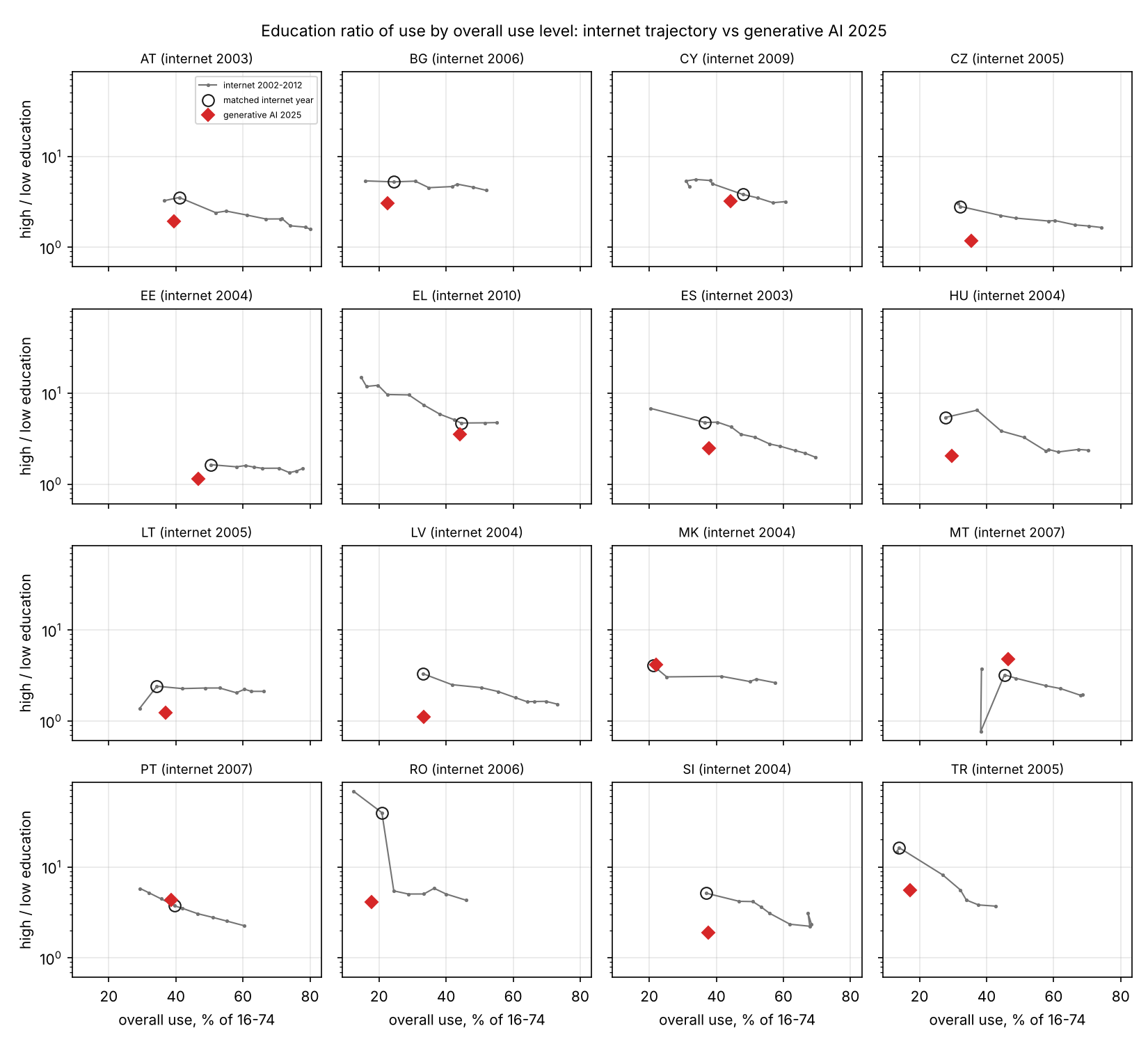}
\caption{Internet trajectory of the high/low education ratio against
overall use, 2002 to 2012, for each matched country; matched year
ringed, 2025 generative AI point overlaid.}
\end{figure}

\begin{figure}
\centering
\includegraphics[width=1\textwidth,height=\textheight]{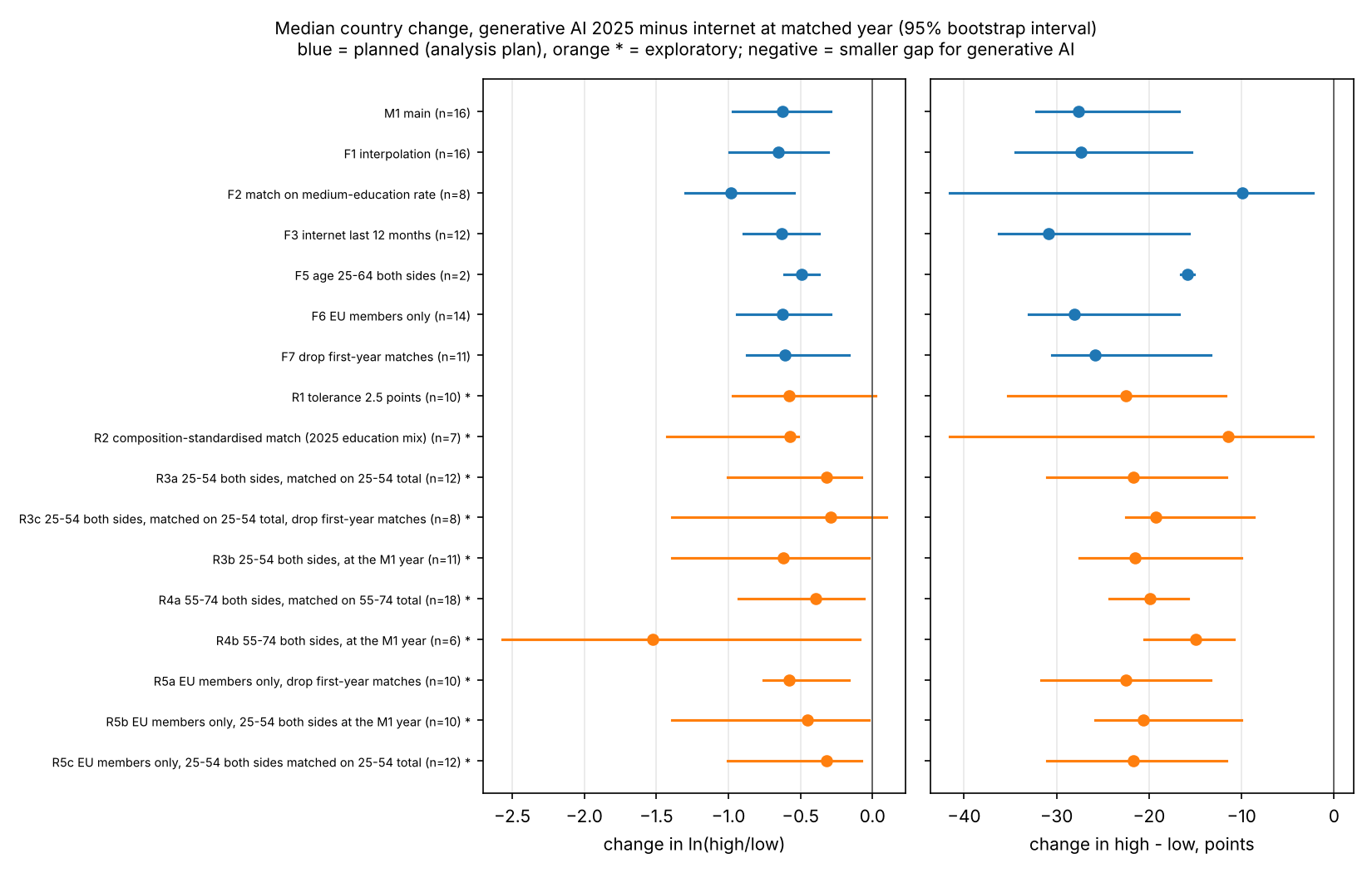}
\caption{Median change and 95\% bootstrap interval for the analyses in
Tables 2 and 3 (except F4 and R6). Blue: planned; orange with asterisk:
exploratory.}
\end{figure}

\hypertarget{discussion}{%
\subsection{5. Discussion}\label{discussion}}

In the 16 European countries where the comparison was possible, people
with low and high completed education differed less in their 2025 use of
generative AI than they had in their internet use in the year when
overall internet use was at a similar level. The difference is large and
consistent in percentage points: the median high-minus-low gap roughly
halves, and the point change is negative in nearly every country in
every generative-AI-versus-internet sensitivity analysis, including
comparisons restricted to the same age band on both sides. (The banking
comparison F4 is a different contrast and is discussed below.) As a
ratio, the difference is smaller in most countries but is not robust: it
disappears under a stricter match tolerance, after removing countries
matched at their first internet observation, and in some age-restricted
variants.

The two measures can diverge because they answer different questions. A
percentage-point gap at a given overall level depends strongly on the
education mix of the population, and Europe's population in 2025 has a
larger high-education share than in the mid-2000s. The composition-aware
match (R2, 7 countries) and the medium-education match (F2, 8 countries)
give point changes of about -11 and -10 points instead of -28. These
matches change the matched years and the set of countries, so the
comparison shows that the size of the point estimate is sensitive to how
composition enters the match; it does not identify what share of the M1
change is due to composition. The ratio is less sensitive to composition
but very sensitive to small low-education rates, which are common in the
internet-era data for some countries (Romania 1.7\% in 2006, Turkey
3.8\% in 2005).

The internet banking comparison (F4) gives no evidence that a technology
which spread after the internet automatically has a smaller education
gap than the internet at a similar level. On the ratio, banking had a
larger gap than the internet in 12 of 15 countries; on points, the
banking and internet gaps could not be distinguished (8 of 15 negative).
This weighs against reading the generative AI result as a mechanical
consequence of being a later technology, but with one comparator and a
null point result it does not rule that reading out.

These findings are in line with the normalization reading of diffusion
in one narrow sense: the recent-use gap for generative AI at roughly one
third of adults is smaller than it was for the internet. They do not
address skills, depth or quality of use, where van Dijk and Hacker
(2003) expected gaps to persist, and they do not address the countries
where the internet spread earliest.

\hypertarget{limitations}{%
\subsection{6. Limitations}\label{limitations}}

\begin{itemize}
\tightlist
\item
  Selection of countries. The matched countries are those whose internet
  series starts below or near their 2025 generative AI rate. Northern
  and western European countries with early internet adoption are
  excluded. Results do not generalise to the EU as a whole.
\item
  Not the same stage of diffusion. The match is on a similar observed
  overall rate among 16 to 74 year olds, not on a common stage of a
  diffusion process. The populations differ in age structure, student
  share, employment and education mix across roughly two decades.
\item
  Unadjusted contrasts. Education is highest level completed; young
  people still in education are counted in lower groups. The comparison
  is a description of unadjusted contrasts, not an education effect
  adjusted for age, student status or employment. Symmetric age-band
  checks (R3, R4) address age only partly.
\item
  One year of generative AI data, a new survey question in 2025, and
  possible differences in translation and understanding across
  countries.
\item
  ISCED 1997 for the internet years and ISCED 2011 for 2025.
\item
  Inference. Eurostat does not publish standard errors in these tables,
  so within-country sampling error is not modelled. The sign tests and
  bootstrap intervals treat countries as units, but the countries are a
  deterministic, data-availability-selected set, not a random sample,
  and neighbouring countries may be correlated. The p-values describe
  how far the signs of 16 point estimates depart from a fair-coin null;
  they are not inference about a wider population of countries.
\item
  Not pre-registered. The analysis plan is not public, so neither its
  timing relative to the computation nor the conformity of the reported
  analyses to it can be verified; the labels ``planned'' and
  ``exploratory'' are the author's classification and every result may
  be read as exploratory. The author had also seen some values and
  expected the direction (Section 3.1).
\item
  Provenance. The extracts are verifiable only as the deposited files,
  their hashes and a same-day byte-identical re-fetch (Section 2); the
  earlier fetches used for the first-pass computation were not recorded,
  and nothing about them is claimed.
\item
  Prior work. The literature search queries before this paper were not
  recorded, the Eurostat (2026b) statistical report, Liu and Wang (2026)
  and Adăscăliței (2026) were found only in later checks, and
  Adăscăliței (2026) was not retrievable by script (Section 9).
\end{itemize}

\hypertarget{conclusion}{%
\subsection{7. Conclusion}\label{conclusion}}

Under the decision rule stated in Section 3.3 (from an analysis plan
that cannot be verified as pre-registered), the primary test finds that
the education gap in generative AI use in 2025 is smaller than the
internet's gap at a similar overall level of use, in 16 European
countries with comparable data. The percentage-point result keeps its
sign in every generative-AI-versus-internet check, but its size is
sensitive to matching that takes education composition into account
(about -10 to -11 points instead of -27.6, in 7 and 8 countries). The
ratio result is fragile. The finding is descriptive, limited to
countries where the internet spread relatively late, and silent on depth
or quality of use.

\hypertarget{changes-from-the-analysis-plan}{%
\subsection{8. Changes from the analysis
plan}\label{changes-from-the-analysis-plan}}

\begin{itemize}
\tightlist
\item
  The author reports that no planned calculation was changed in a way
  that alters a result; because the plan is not public, this cannot be
  checked by readers. All planned analyses (M1, F1 to F7) and stop
  conditions are reported. No stop condition prevented the primary
  statement. The F5 condition applied in its fewer-than-8 branch
  (Section 4.3), so the age-matched comparison at ages 25 to 64 is
  reported as not possible with these data.
\item
  Flag rule (added after the first results, so an exploratory
  amendment): every cell used, including the medium-education cells in
  F2, the student and 16-24 cells, and the LFS education cells in R2, is
  treated as missing when it carries a u (low reliability) flag, and the
  R2 education weights come from one year for all three groups. Rerun on
  the same data, the summary and per-country output files are
  byte-identical to the earlier run. The only change is the student
  comparison (one u-flagged cell, North Macedonia's 2025 generative AI
  rate among students, is dropped; median 85.5\% over 15 countries
  instead of 83.5\% over 16). Later changes left every analysis value
  unchanged: the column \texttt{preregistered} in
  \texttt{out/idea3\_rev\_summary.csv} was renamed \texttt{planned} (and
  the key \texttt{prereg} in the results JSON to \texttt{planned}); the
  results JSON now also holds the per-country F4 values;
  \texttt{audit\_flags.py} was extended to list all cells entering M1,
  F2, R2 and the student and 16-24 comparison, including the R2
  weight-year rule, and to write
  \texttt{out/flag\_audit\_candidates.csv}, the cells examined by the
  matching step of M1, F2 and R2 in years that were not selected
  (Section 4.3); and \texttt{describe\_m1.py} (descriptive medians after
  Table 1) and \texttt{verify\_extracts.py} (same-day re-fetch
  comparison, Section 2) were added. After these changes,
  \texttt{analysis.py}, rerun on the same data, gave byte-identical
  output files.
\item
  The planned analysis was originally run with a first-pass script that
  is not public. The analysis script in the ancillary files
  (\texttt{anc/analysis.py}, an earlier revision script plus the flag
  rule above) re-implements M1 and F1 to F7 with, according to the
  author, the same rules (the first-pass script and the plan are not
  public) and adds the exploratory analyses R1 to R6. Its outputs
  reproduce the earlier revision's output files byte for byte from data
  fetched again on 2026-10-26.
\item
  The analysis plan specified figure content; the figures here follow it
  (Figure 1) and add a summary forest plot (Figure 2).
\item
  All analyses labelled R, the student-rate comparison and the
  asymmetric 25-64 comparison were added after the primary result was
  known.
\end{itemize}

\hypertarget{literature-search}{%
\subsection{9. Literature search}\label{literature-search}}

Searches on 2026-10-26 (web search, three queries on generative AI
adoption versus internet diffusion by education in Europe; digital
divide normalization and stratification; ChatGPT use by education level)
and the searches noted in the analysis plan (eight searches on
2026-10-26) found no study comparing the education gap in generative AI
use with the internet's at a matched level of use. The exact query
strings and the screening of results were not recorded for these
searches, so the absence of such a study is not established. Later
checks of the manuscript identified four relevant sources that these
searches missed: Suárez and García-Mariñoso (2025), Adăscăliței (2026),
the Eurostat (2026b) statistical report and Liu and Wang (2026). The
Suárez and García-Mariñoso record was read on ideas.repec.org. The
Eurostat (2026b) record page and PDF were downloaded from ec.europa.eu
on 2026-10-26 and the generative AI section (Section 3.2, pp.~38 onward)
was read, including the education figures quoted in Section 1. The
Adăscăliței (2026) working paper could not be retrieved by script (on
2026-10-26 its publication page and file link answered HTTP 429); the
author read its full text and confirmed on 2026-10-27 the details used
in Section 1 and in the reference list. It is a descriptive analysis of
Eurostat aggregate data (isoc\_ai\_iaiu) for the EU27 in 2025, with no
multivariate analysis. It calls education a strong predictor of use and
notes a threshold at tertiary education. Its only comparison with
earlier technology waves concerns the gender gap, which it finds smaller
than historical gaps, and a remark that the country ranking partly
replicates earlier waves; it does not compare the education gap with the
internet's or that of other earlier technologies. The bibliographic
record of Liu and Wang (2026) (title, authors, World Development 199,
article 107260, March 2026) was checked against Crossref on 2026-10-26.
The publisher page answered HTTP 403, so the journal version was not
read; the description in Section 1 is from the earlier World Bank Policy
Research Working Paper 10870 (August 2024), downloaded from
documents.worldbank.org on 2026-10-26 and read in its abstract,
introduction, user-demographics section and conclusion. The journal
version may differ. Bibliographic details of the journal articles below
were checked against Crossref or the publisher or repository page, and
those of the two books against Crossref (Norris) and Open Library
(Rogers), on 2026-10-26.

\hypertarget{references}{%
\subsection{References}\label{references}}

Adăscăliței, D. (2026). Who uses Generative AI? Patterns and
inequalities across the EU. Eurofound working paper WPEF26003, Dublin,
27 February 2026. Not subject to the full Eurofound evaluation,
editorial and publication process.
\url{https://www.eurofound.europa.eu/en/publications/all/who-uses-generative-ai-patterns-and-inequalities-across-the-eu}

Bick, A., Blandin, A., and Deming, D. J. (2024, revised 2025). The Rapid
Adoption of Generative AI. NBER Working Paper 32966.
\url{https://www.nber.org/papers/w32966}. Published as: Management
Science 72(10), 8440-8456 (2026).
\url{https://doi.org/10.1287/mnsc.2025.02523}

Daepp, M. I. G., and Counts, S. (2025). The Emerging Generative
Artificial Intelligence Divide in the United States. Proceedings of
ICWSM 2025. arXiv:2404.11988. \url{https://arxiv.org/abs/2404.11988}

Eurostat (2025). 32.7\% of EU people used generative AI tools in 2025.
Eurostat news, 16 December 2025.
\url{https://ec.europa.eu/eurostat/web/products-eurostat-news/w/ddn-20251216-3}

Eurostat (2026a). 64\% of 16-24-year-olds used AI in 2025. Eurostat
news, 10 February 2026.
\url{https://ec.europa.eu/eurostat/web/products-eurostat-news/w/edn-20260210-1}

Eurostat (2026b). The use of artificial intelligence (AI) technologies
in the European Union: Key results, 2026 edition. Statistical report
KS-01-26-009, 26 March 2026. Publications Office of the European Union.
\url{https://doi.org/10.2785/9221093}

Eurostat. Use of generative AI tools by individuals (isoc\_ai\_iaiu),
last updated 5 June 2026. \url{https://doi.org/10.2908/isoc_ai_iaiu}

Eurostat. Individuals - internet use (isoc\_ci\_ifp\_iu), last updated
17 April 2026. \url{https://doi.org/10.2908/isoc_ci_ifp_iu}

Eurostat. Individuals - internet activities (isoc\_ci\_ac\_i), last
updated 17 April 2026. \url{https://doi.org/10.2908/isoc_ci_ac_i}

Eurostat. Population by educational attainment level, sex and age
(lfsa\_pgaed), last updated 10 September 2026.
\url{https://doi.org/10.2908/lfsa_pgaed}

Eurostat. ICT usage in households and by individuals, reference metadata
(isoc\_i\_esms).
\url{https://ec.europa.eu/eurostat/cache/metadata/en/isoc_i_esms.htm}

Eurostat. Copyright notice and free re-use of data.
\url{https://ec.europa.eu/eurostat/help/copyright-notice}

Galácz, A., and Smahel, D. (2007). Information Society from a
Comparative Perspective: Digital Divide and Social Effects of the
Internet. Cyberpsychology: Journal of Psychosocial Research on
Cyberspace 1(1). \url{https://cyberpsychology.eu/article/view/4204}

Humlum, A., and Vestergaard, E. (2025). The unequal adoption of ChatGPT
exacerbates existing inequalities among workers. Proceedings of the
National Academy of Sciences 122(1), e2414972121.
\url{https://doi.org/10.1073/pnas.2414972121}

Liu, Y., and Wang, H. (2026). Who on earth is using Generative AI? World
Development 199, 107260.
\url{https://doi.org/10.1016/j.worlddev.2025.107260}. Earlier version:
World Bank Policy Research Working Paper 10870 (2024).
\url{https://documents.worldbank.org/en/publication/documents-reports/documentdetail/099720008192430535}

Norris, P. (2001). Digital Divide: Civic Engagement, Information
Poverty, and the Internet Worldwide. Cambridge University Press. ISBN
978-0-521-80751-7 (hardback), 978-0-521-00223-3 (paperback).
\url{https://doi.org/10.1017/CBO9781139164887}

Rogers, E. M. (2003). Diffusion of Innovations, 5th ed.~Free Press. ISBN
978-0-7432-2209-9. Chapter 7.

Suárez, D., and García-Mariñoso, B. (2025). On the verge of a digital
divide in the use of generative AI? Telecommunications Policy 49(7),
102997. \url{https://doi.org/10.1016/j.telpol.2025.102997}

van Dijk, J., and Hacker, K. (2003). The Digital Divide as a Complex and
Dynamic Phenomenon. The Information Society 19(4), 315-326.
\url{https://doi.org/10.1080/01972240309487}

\hypertarget{use-of-ai-tools}{%
\subsection{Use of AI tools}\label{use-of-ai-tools}}

Claude (Anthropic, claude-opus-5-5) wrote the data-retrieval and
analysis code and drafted the manuscript and figures. Codex (OpenAI)
reviewed the manuscript and code over six rounds. The author reviewed
the manuscript, checked the cited references against their sources, and
is responsible for the content.

\end{document}